\documentclass[a4paper,fleqn,usenatbib]{mnras}

\usepackage[T1]{fontenc}
\usepackage{ae,aecompl}

\usepackage{graphicx}	
\usepackage{amsmath}	
\usepackage{amssymb}	
\usepackage{subfig}
\usepackage{hyperref}

\title[Flux and Photon Index distributions of blazars]{Study of long-term flux and photon  Index distributions of blazars using RXTE observations}

\author[R. Khatoon et al.]{
Rukaiya Khatoon$^{1}$\thanks{E-mail: rukaiyakhatoon12@gmail.com},
Zahir Shah$^{2}$\thanks{zahir@iucaa.in}, Ranjeev Misra$^{2}$  and Rupjyoti Gogoi$^{1}$  
\\
$^{1}$Tezpur University,Napaam-784028, Assam, India.\\
$^{2}$Inter-University Center for Astronomy and Astrophysics, Post Bag 4, Ganeshkhind, Pune-411007, India. \\
}

\date{Accepted XXX. Received YYY; in original form ZZZ}

\pubyear{2019}

\begin{document}
\label{firstpage}
\pagerange{\pageref{firstpage}--\pageref{lastpage}}
\maketitle

\begin{abstract}

We present a detailed study of flux and index distributions of three blazars (one FSRQ and two BL\,Lacs) by using 16 years of Rossi X-ray Timing Explorer archival data. The three blazars were chosen such that their flux and index distributions have sufficient number of data points ($\geq$90) with relatively less uncertainty ($\overline{\sigma_{err}^{2}}/\sigma^{2}$<0.2) in light curves. Anderson-Darling (AD) test and histogram fitting shows that flux distribution of FSRQ 3C273 is log-normal, while its photon index distribution is Gaussian. This result is consistent with linear Gaussian perturbation in the particle acceleration time-scale, which produces log-normal distribution in flux. However, for two BL\,Lacs viz. Mkn501 and Mkn421, AD test shows that their flux distributions are neither Gaussian nor log-normal, and their index distributions are non-normal. The histogram fitting of Mkn501 and Mkn421, suggests that their flux distributions are more likely to be a bi-modal, and their index distributions are double Gaussian. Since, Sinha et al. (2018) had shown that Gaussian distribution of index produces a log-normal distribution in flux, double Gaussian distribution of index in Mkn501 and Mkn421 indicates that their flux distributions are probably double log-normal. Observation of double log-normal flux distribution with double Gaussian distribution in index reaffirms two flux states hypothesis. Further, the difference observed in the flux distribution of FSRQ (3C273) and BL Lacs (Mkn501 and Mkn421) at X-rays, suggest that the low energy emitting electrons have a single log-normal flux distribution while the high energy ones have a double log-normal flux distribution.\\

\end{abstract}



\section{Introduction}

 Blazars are the class of radio loud active galactic nuclei (AGNs), with relativistic jets oriented close to the observer's line of sight \citep{1995PASP..107..803U}. The non-thermal emission from blazars extends from radio to $\gamma$-ray energies. Blazars are characterized by their extreme properties like luminous core, rapid variability, high polarization, superluminal motion, occasional spectacular flares etc. These extreme properties are usually attributed to the Doppler boosting of the emission which occurs from the relativistic regions in the jet.
Blazars are broadly classified into two sub-classes, namely, BL\,Lac objects and Flat Spectrum Radio Quasars (FSRQs). The difference between these two sub-classes is based on the presence or absence of emission/absorption line features in their optical spectrum, such that FSRQs show strong emission lines whereas for BL\,Lacs, the emission lines are weak/absent \citep{1995PASP..107..803U, 2003ApJ...585L..23F}.
The non-thermal emission originating from blazar jet produces a double humped spectral energy distribution (SED) \citep{1998MNRAS.299..433F}, with the low energy component peaking at optical/UV/soft X-ray energies, while the high energy component peaks at GeV energies. The low energy component is well established to be produced by synchrotron emission from relativistic electrons gyrating in the magnetic field of the jet, whereas the high energy component is usually attributed either to the inverse Compton (IC) scattering of low energy photons \citep{1992ApJ...397L...5M, 1992A&A...256L..27D, 2017MNRAS.470.3283S} or to the hadronic cascades initiated in the jet \citep{2007Ap&SS.307...69B}. Based on the location of the peak frequency in the low energy component, blazars are further classified into three sub-classes namely, high energy peaked BL\,Lac (HBL; $\nu_{p}>10^{15.3}$ Hz; \citep{1995ApJ...444..567P}), intermediate energy peaked BL\,Lac (IBL; $10^{14} < \nu_{p} {\leq} 10^{15.3}$ Hz), and low energy peaked BL\,Lac (LBL; $\nu_{p} {\leq} 10^{14}$ Hz) \citep{2016ApJS..226...20F}.
 
The light curves of blazars show unpredictable variations over a broad range of time-scales ranging from minutes to years across the entire electromagnetic spectrum. The clue to the mechanism causing such variations can be obtained by studying their long-term flux distributions. Typically, a Gaussian distribution of fluxes suggests additive processes, which is obtained when the flux variation is stochastic and linear. However, if the stochastic flux variation is non-linear, a Gaussian distribution in logarithmic flux values is expected, and such distributions are known as log-normal, which are often found in galactic and extra-galactic sources, like X-ray binaries, gamma-ray bursts, and AGNs \citep{lognorm_xrb, 2002PASJ...54L..69N, 1997MNRAS.292..679L, 2002A&A...385..377Q, 2009A&A...503..797G}. In the case of AGNs, the log-normal behavior are observed on time-scales ranging from minutes to days \citep{2004ApJ...612L..21G}; whereas, for X-ray binaries, such behavior is seen in sub-second time scales \citep{2005MNRAS.359..345U}. Among blazars, BL\,Lac is the first blazar in which log-normal distribution of X-ray flux was observed  \citep{2009A&A...503..797G}. Subsequently, such log-normal behavior in fluxes with the excess r.m.s correlating linearly with the average flux, have been found in different energy-bands for a HBL, PKS\,2155-304 \citep{2017A&A...598A..39H}.  Recently, a log-normal distribution of flux was shown in PKS\,2155-304 at X-ray and optical bands by using ten years of data \citep{2019MNRAS.484..749C}.  
In the very high energy (VHE) $\gamma$-ray band, the log-normal behavior of flux distribution has been detected in the well known high synchrotron peaked BL\,Lac objects Mrk\,501 \citep{2010A&A...524A..48T, 2018Galax...6..135R} and  PKS 2155-304 \citep{2009A&A...502..749A, 2010A&A...520A..83H}. 
The brightest blazar sources seen by \emph{Fermi}-LAT show a similar trend in their long term monthly binned $\gamma$-ray light curves \citep{2018RAA....18..141S}. Also, using the \emph{Fermi}-LAT observations, the long-term $\gamma$-ray flux variability of the VHE source 1ES 1011+496 is independently confirmed to be log-normal by \citealp{my1011}. In addition to single log-normal distribution, the double log-normal profile has also been revealed at multi-wavelength bands for some blazars  \citep{pankaj_ln, 2018RAA....18..141S}.

 The log-normal behavior of these astrophysical sources is generally explained in terms of variations from the accretion disk, which are multiplicative in nature \citep{2005MNRAS.359..345U, 2010LNP...794..203M}.  However, the fluctuations in the accretion disk may not produce minute scale variability as seen in most of the blazar sources \citep{1996Natur.383..319G, vhe501, vaidehi421}. Therefore, as an alternative to the accretion disk model, the log-normal flux distribution in blazars have also been explained in terms of a sum of emission from a large collection of mini-jets, which are randomly oriented inside the relativistic jet \citep{minijet}. Moreover, recently it has been shown that a linear Gaussian perturbation in the particle acceleration time scale can produce the log-normal flux distribution \citep{2018MNRAS.480L.116S}. They showed that perturbation in the acceleration time scale produces a Gaussian distribution in the index which in turn results in a log-normal distribution of the flux, whereas the perturbation in the particle cooling rate produces neither a Gaussian nor a Log-normal flux distribution.

 In this work, we study the flux and photon index distribution properties of blazar sources having statistically significant light curves in the 16 years of RXTE observation. The sample selection criteria from the RXTE AGN catalog is described in Section \$2, Characterization of flux/index distribution and correlation study between the flux and photon index is described in Section \$3. The physical interpretations of the obtained flux distributions are discussed in Section \$4.

\section{RXTE archieve and Sample selection}

 The Rossi X-ray Timing Explorer (RXTE) public database provides the systematically analyzed long-term light curves and spectral information of AGN sources, for the period from January 1996 to January 2012. RXTE has two co-aligned instruments, the Proportional Counter Array (PCA; 2-60 keV) \citep{2006ApJS..163..401J} and the High-Energy X-ray Timing Experiment (HEXTE; 15-250 keV) \citep{1998ApJ...496..538R}. It provides light curves in the energy range 2--10 keV.  In these light curves, the sampling of data is uneven as different periods had been proposed for various scientific goals.  However, despite the time gaps in the light curves, the sixteen years of  RXTE observation provides a large data set, which are some times statistically suitable to investigate the variability behavior.  

\begin{figure}
\centering
\includegraphics[scale=0.44,angle=-90]{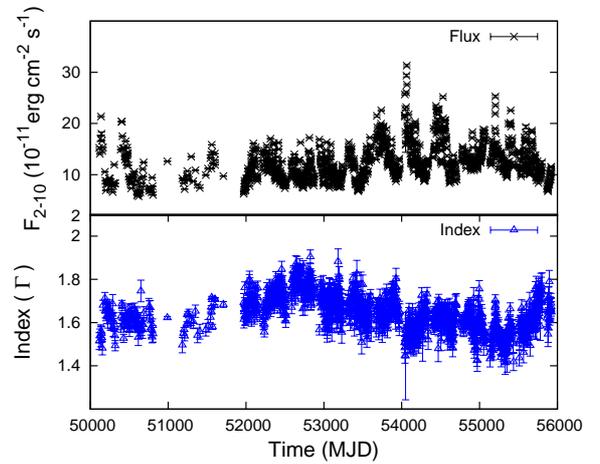}
\vspace{-0.4cm}
\caption{The X-ray flux/index light curves of 3C\,273 obtained by using the sixteen years of RXTE archive data. Top panel: 2-days time binned flux light curve obtained in the energy range 2-10\,keV, bottom panel: 2-days time binned index light curve.}

\label{fig:3c279}
\end{figure}

\begin{figure}
\vspace{-0.4cm}
\centering
\includegraphics[scale=0.44,angle=-90]{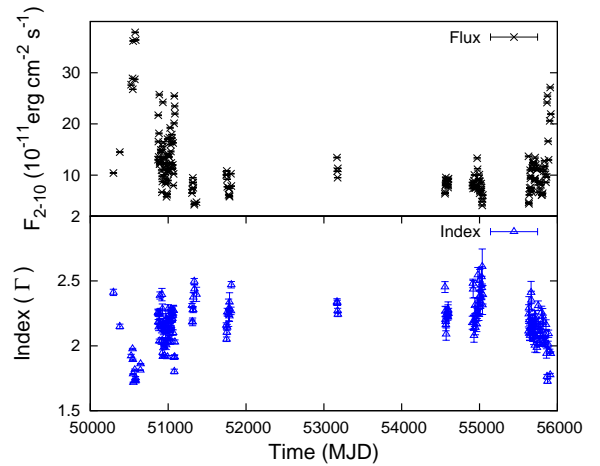}
\vspace{-0.4cm}
\caption{The X-ray flux and $\Gamma1$ light curves of  Mkn\,501 obtained by using the sixteen years of RXTE archive data. Top panel and bottom panel are the same as in Fig. \ref{fig:3c279}.}

\label{fig:mkn501}
\end{figure}

\begin{figure}
\vspace{-0.4cm}
\centering
\includegraphics[scale=0.44,angle=-90]{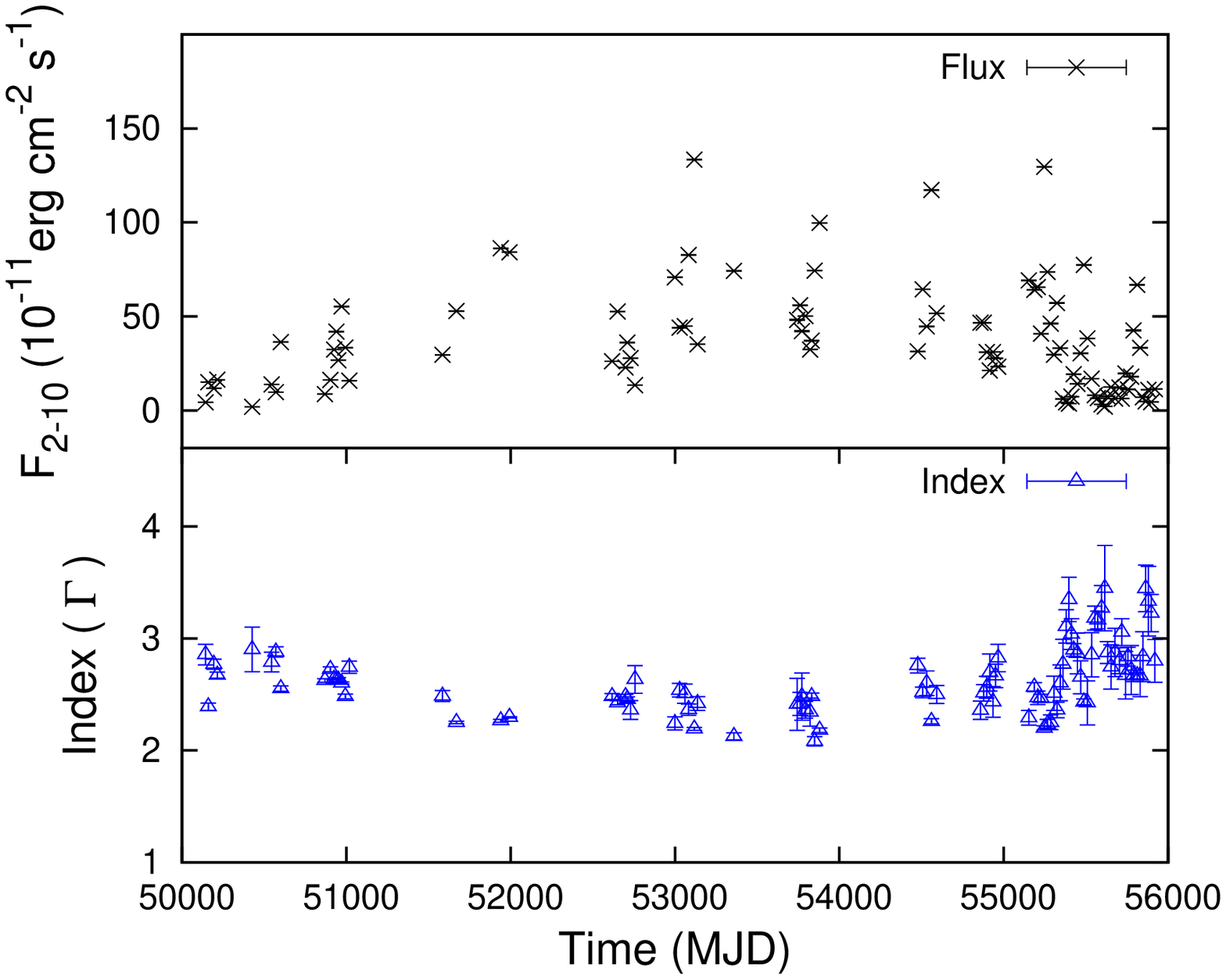}
\vspace{-0.4cm}
\caption{The X-ray flux and $\Gamma1$ light curves of Mkn\,421 obtained by using the sixteen years of RXTE archive data. Top panel: 10-days time binned flux light curve obtained in the energy range 2-10\,keV, bottom panel: 10-days time binned $\Gamma1$ light curve.}

\label{fig:mkn421}
\end{figure}

 We first selected all the blazars from the RXTE AGN Timing \& Spectral Database (RATSD \footnote{\url{https://cass.ucsd.edu/~rxteagn/}}), for which the flux and photon spectral index light curves are available with more than 90 number of flux/index points. The spectra of these blazars in RATSD are mostly fitted with simple power-law model, except few BL Lac sources which are fitted better with a broken power-law model with break energy below $\approx$ 10 keV \citep{2013ApJ...772..114R}.  The RATSD provide light curves in the energy range 2-10 keV, 2-4 keV, 4-7 keV and 7-10 keV.   However, in order to have flux light curves with good statistics,  we consider flux light curves which have been obtained in the full energy range of  2--10 keV. Further, to select the light curves with good statistics, we define light curve significance fraction as 

\begin{equation}
R=\frac{\overline{\sigma_{err}^2}}{\sigma^2}
\end{equation}
where $\overline{\sigma_{err}^2}$ is the mean square error of flux/index distribution and $\sigma^2$ is the variance of flux/index distribution.  From the selected blazars, only those sources are considered for further analysis which has $R<0.2$ in both the flux and index light curves. However, this condition was not satisfied in the unbinned index light curves (light curves taken directly from RATSD) of the blazars except Mkn\,501, therefore, for each selected blazar, we binned the flux/index light curves by combining the flux/index points from 2 days  to a maximum of 10 days. The upper limit of 10 days bin is chosen to ensure that the minimum number of flux/index points in the binned light curve is $\geq90$.  
After implementing these conditions i.e., length of the light curve $\geq90$ and $R<0.2$, we are restricted to two BL\,Lacs viz. Mkn\,501, Mkn\,421 and one FSRQ viz. 3C\,273. The above conditions are met with  2-days time-bin in 3C\,273 and 10-days time-bin for Mkn 421, while in Mkn\,501, these conditions are satisfied in unbinned and two-days time binned light curves (see Table \ref{tab:R}). However, in order to obtain evenly sampled light curves in all three sources, we have used two-days time binned light curve for Mkn\,501. The length of light curves and the `R' values of 3C\,273, Mkn\,501 and Mkn\,421 are given in Table \ref{tab:R}. \\
3C\,273 is a well studied FSRQ source located at a redshift $z\sim0.158$. The RXTE X-ray spectra of this source is fitted with a power-law model with an average power-law index of $\Gamma=$ 1.70$\pm$0.01 \citep{2013ApJ...772..114R}. The obtained RXTE flux/index light curves of 3C\,273 are shown in Fig. \ref{fig:3c279}. The two BL Lac objects viz. Mkn\,501 and Mkn\,421 at a redshift of 0.033 and 0.031 respectively, are very well known high synchrotron peaked BL Lac (HBL) objects, with the low energy SED component peaking at frequency $\nu_{p}> 10^{15.3}$ Hz. The RXTE  spectra of these sources are fitted better with a broken power-law model \citep{2013ApJ...772..114R}. For Mkn\,501, the average values of power-law photon index before break energy ($\Gamma1$) and after break energy ($\Gamma2$) are obtained as 1.97$\pm$0.02 and 2.02$\pm$0.01 respectively with a break energy at 6.9$\pm$1.2 keV, where as for Mkn\,421, the average values of  $\Gamma1$ and  $\Gamma2$ are obtained as 2.41$\pm$0.09 and 2.75$\pm$0.01 with break energy of 6.6$\pm$0.4 keV. The RXTE flux and $\Gamma1$ light curves of Mkn\,501 and Mkn\,421 are shown in Figs \ref{fig:mkn501} and \ref{fig:mkn421}.

Moreover, since RATSD provides light curves from a pointing instrument (PCA), there is a possibility of bias being introduced in the flux distributions, if the sources were preferentially observed in the high flux states. To check for this, we looked for correlation between the flux and time-gap (time between two observations) with the Spearman's rank correlation test. The null hypothesis probability values (P-value) for 3C273, Mkn 501 and Mkn 421 are obtained as 0.07, 0.15 and 0.08 respectively. In the case of Spearman's correlation, the null hypothesis ($H_{0}$) is that there is no correlation between two variables, and we reject the null hypothesis if P<0.01. Therefore, the obtained P-values indicate that there is no significant correlation between flux and time-gap.

\begin{table}
\centering
\caption{Light curve significance fraction R values for the unbinned/binned flux and index light curves. Col:- \,1: Selected blazars satisfying the conditions R<0.2 and length of binned flux/index light curve $\geq 90$, 2: Number of data points in the distributions, 3: R-value for index light curve, and 4: R-value for the flux light curve.}
\begin{tabular}{lccc}
\hline
\hline
Blazar name   &  \multicolumn{1}{c}{Number of }    &  \multicolumn{1}{c}{$R_\Gamma$} &   \multicolumn{1}{c}{$R_{Flux}$}\\ 
& data points & & \\   
							   \hline

3C\,273   & 1960 (unbinned)  & 0.22 & $2.0\times10^{-3}$\\

  & 1151 (2-days binned) & 0.19 & $1.5\times10^{-3}$\\               
  \hline
  
Mkn\,501   & 496 (unbinned) & 0.05  & $1.0\times10^{-4}$\\

 & 188 (2-days binned) & 0.06 & $1.0\times10^{-4}$\\
 \hline
 Mkn\,421 & 1182 (unbinned) & 0.75  & $6.8\times10^{-5}$\\
 & 93 (10-days binned) & 0.17 & $9.6\times10^{-6}$\\

	 \hline    
    
        \end{tabular}
       \label{tab:R}
        						   
\end{table}  

\section{Distribution study for blazars}

 \subsection{AD test} \label{sec:tau_a_pert}
Generally, the flux distribution of blazar light curves shows an asymmetric/tailed trend. Anderson-Darling (AD) test statistic is an useful tool for the normality test, which is sensitive towards the tails of a distribution \citep{1992nrfa.book.....P}. The null hypothesis ($H_0$) of the AD test is that the data sample is drawn from a particular distribution (say normal distribution in our case). The AD test calculates the null hypothesis probability value (p--value) such that p--value < 0.01 would indicate the deviation from the normality of the sample. 

The AD test for 3C\,273 shows that the p--values for flux in log-scale and index in normal-scale are larger than 0.01, which suggests the flux distribution is consistent with a log-normal and the index distribution with a normal one. However, for the two  BL\,Lac objects viz. Mkn\,501 and Mkn\,421, p--values for flux in linear-scale and log-scale are much smaller than 0.01, which indicates that the flux distribution will be neither normal nor log-normal. Moreover, the p--value of index distribution also suggests a non-normal distribution of index in Mkn\,501 and Mkn\,421. The AD test results of 3C\,273, Mkn\,501 and Mkn\,421 are summarized in Table \ref{tab:AD}.


\subsection{Histogram of Flux and Index}
Histogram fitting is also a helpful tool to characterize the nature of the distribution. Here, we construct the normalized histograms of the logarithm of flux and photon index in linear-scale for the three selected blazars. The bin-widths of the histogram are chosen such that each bin carries an equal number of data points. In the case of FSRQ 3C\,273, the single peak in the flux and index histograms suggest for the single distribution; hence we fitted these histograms with the probability density function (PDF) given by
\begin{equation}\label{eq:gauss}
\rm f(x) = \frac{1}{\sqrt{2\pi \sigma^2}} e^\frac{-(x-\mu)^2}{2\sigma^2}
\end{equation}

where $\rm \mu$  and $\rm \sigma$ are the centroid and width of the distribution.  
The flux and index histograms along with best fitted PDF (equation \ref{eq:gauss}) are shown in the multiplot (Fig. \ref{fig:273_corr}) and the best fit parameters are summarized in Table \ref{tab:2distpar_fsrq}. The fit parameters confirm that the flux distribution of 3C\,273 is log-normal while its index is normally distributed. These results are consistent with the results obtained from the AD test statistic. On the other hand, the flux and index histograms of the two BL\,Lacs (Mkn\,501 and Mkn\,421) show a double-peaked structure. Also, the AD test results of these two BL\,Lacs show that the flux distribution is neither Gaussian nor log-normal, and the index distribution is not Gaussian. We therefore analyzed these distributions by fitting their histograms with double PDF

\begin{equation}\label{eq:dpdf}
 \rm d(x) = \frac{a}{\sqrt{2\pi \sigma_1^2}} e^\frac{-(x-\mu_1)^2}{2\sigma_1^2} \
       + \frac{(1-a)}{\sqrt{2\pi \sigma_2^2}} e^\frac{-(x-\mu_2)^2}{2\sigma_2^2} 
\end{equation}
where, $\rm a$ is the normalization fraction, $\rm \mu_1$ and $\rm \mu_2$ are the centroids of
the distribution with widths $\rm \sigma_1$ and $\rm \sigma_2$, respectively. The fit of flux histogram in log-scale and index histogram in linear-scale with equation \ref{eq:dpdf} will result in the double log-normal fit of the flux distribution and double normal fit of the index distribution. During the double log-normal fit of flux histogram, we have kept all parameters free viz. $\mu_1$, $\mu_2$, $\sigma_1$,  $\sigma_2$ and a. However, in case of double normal fit of index histogram, we have fixed the normalization fraction `a' to the best fit parameter value obtained in the double log-normal flux histogram fit and carried fitting with free parameters as $\mu_1$, $\mu_2$, $\sigma_1$ and $\sigma_2$.  The same normalization fraction in the flux and index double PDF will ensure the similar contribution of the respective components in the flux and index distribution. The best fit parameter values of fitting the double distribution flux/index histograms with equation \ref{eq:dpdf}  are given in Table \ref{tab:2distpar} and corresponding plots are shown in Figs \ref{fig:501_corr} and \ref{fig:421_corr}. In case of Mkn\,421, the error on the normalization fraction `a' is not well constrained due to poor data statistics in the flux histogram, its best fit value ranges from 0.05--0.95. In this case, we have fixed the normalization fraction `a' for both flux and index distribution as 0.3 (Table \ref{tab:2distpar}).  The double log-normal fit to flux histograms of  Mkn\,501 and Mkn\,421 gave a ${\chi^2}/{dof}$ of  ${38.28}/{33}$ and ${11.85}/{15}$ respectively, while the other double distribution functions, such as  combination of log-normal and Gaussian gave a ${\chi^2}/{dof}$ of ${38.38}/{33}$ for Mkn\,501 and ${12.75}/{15}$ for Mkn\,421. Combination of  Gaussian and log-normal gave a ${\chi^2}/{dof}$ of ${50.36}/{33}$ for Mkn\,501 and  ${14.10}/{15}$ for Mkn\,421, while a double Gaussian fit gave a ${\chi^2}/{dof}$ of ${41.58}/{33}$ for Mkn\,501 and ${14.70}/{15}$ for Mkn\,421. These reduced $\chi^2$ values suggest that the double log-normal fit and lognormal+Gaussian fit to the flux histograms of two BL\,Lacs are equally good. The best fit parameter values obtained by fitting the  flux histogram with lognormal+Gaussian PDF are given in Table \ref{tab:2distpar_lg}. Further, we found that the photon index distribution in both Mkn\,501 and Mkn\,421 are fitted with the double Gaussian distribution function with ${\chi^2}/{dof}$ of ${39.1}/{34}$ and $16.94/14$ respectively.

\begin{table*}
		\caption{ AD test results for the flux/index distribution of three selected blazars viz. 3C\,273, Mkn\,501 and Mkn\,421 Col:- 1: Selected
blazars satisfying the conditions R<0.2 and length of binned flux/index light curve $\geq90$, 2: Number of data points in the distributions, 3,4:  AD statistics for Flux and Logarithm of flux distribution, and 5: AD statistics for index distribution.}
        \begin{tabular}{c c c c c}
                \hline
                \hline
				Blazar name  &  \multicolumn{1}{c}{Number of }   &  \multicolumn{1}{c}{Normal (Flux)}            &     \multicolumn{1}{c}{Log-normal (Flux)} & \multicolumn{1}{c}{Normal (Spectral index)}\\ 
				& data points & AD(p--value) & AD(p--value) & AD(p--value) \\ \hline  

3C\,273  & 1151 (2-days binned)  & 13.02 (<$2.2\times10^{-16}$) & 0.76 (0.06) & 0.57 (0.14) \\
\hline
Mkn\,501 & 188 (2-days binned) & 15.89 (<$2.2\times10^{-16}$) & 2.78 ($4.96\times10^{-7}$) & 1.24 ($3.0\times10^{-3}$)\\							   
\hline

Mkn\,421 & 93 (10-days binned) & 2.29 ($7.44\times10^{-6}$) & 1.27 ($2.5\times10^{-3}$) & 1.09 ($7.0\times10^{-3}$) \\

\hline

  	                \hline    
     
        \end{tabular}
        \label{tab:AD}
        
\end{table*}
 
 							    \begin{table*}
		\caption{Best fit parameter values of the PDF (equation \ref{eq:gauss}) fitted to the logarithm of flux and index histograms. Col:- 2: Histogram obtained from the logarithm of flux and linear index distribution,  3,4: Best fit values of $\mu$ and $\sigma$, 5: Degrees of freedom and 6: Reduced $\chi^{2}$.}
        \begin{tabular}{c c c c c c }
                \hline
                \hline
				Blazar name  &  Histogram  & $\mu$ & $\sigma$ &     dof & $\rm \chi^2/dof$ \\

							   \hline

3C\,273 & log10(Flux) & -9.922$\pm$0.004 & 0.124$\pm$0.003 & 21 & 0.92\\

&Index &  1.642$\pm$0.003 & 0.088$\pm$0.002 & 21 & 1.22\\
\hline
\end{tabular}
 \label{tab:2distpar_fsrq}
\end{table*}

					   \begin{table*}
		\caption{Best fit parameter values of the double PDF (equation \ref{eq:dpdf}) fitted to the logarithm of flux and index histograms. Col:- 2: Histogram obtained from the logarithm of flux and linear index distribution,  3--6: Best fit values of $\mu_1$, $\sigma_1$, $\mu_2$ and $\sigma_2$, 7: Normalization fraction,  8: Degrees of freedom  and 9: Reduced $\chi^{2}$. }

\begin{tabular}{c c c c c c c c c}

\hline
  \hline
				Blazar name & Histogram  & $\mu_1$ & $\sigma_1$  & $\mu_2$ & $\sigma_2$  &     \multicolumn{1}{c}{a}   &   dof & $\rm \chi^2/dof$ \\
\hline

Mkn\,501  & log10(Flux) & -9.62$\pm$0.04 & 0.10$\pm$0.03  & -10.02$\pm$0.02 & 0.14$\pm$0.02 & 0.83$\pm$0.06 & 33 & 1.16\\
& Index  & 1.74$\pm$0.03 & 0.09$\pm$0.02 & 2.19$\pm$0.02 & 0.15$\pm$0.01 & 0.83  & 34 & 1.15\\
\hline

Mkn\,421  & log10(Flux)  & -9.36$\pm$0.05 & 0.26$\pm$0.05 & -10.10$\pm$0.08 & 0.29$\pm$0.08 & 0.3 & 15 & 0.79\\
& Index & 2.54$\pm$0.05 & 0.21$\pm$0.04 & 3.09$\pm$0.48 & 0.51$\pm$0.31 & 0.3 & 14 & 1.21\\

\hline

        \end{tabular}
        
 \label{tab:2distpar}
        
\end{table*}

 \begin{table*}
		\caption{Best fit parameter values of the double PDF (lognormal+Gaussian) fitted to the logarithm of flux histogram. Col:- 2: Histogram obtained from the logarithm of flux distribution,  3--6: Best fit values of $\mu_1$, $\sigma_1$, $\mu_2$ and $\sigma_2$, 7: Normalization fraction,  8: Degrees of freedom  and 9: Reduced $\chi^{2}$.}

\begin{tabular}{c c c c c c c c c}

\hline
  \hline
				Blazar name & Histogram  & $\mu_1$ & $\sigma_1$  & $\mu_2$ & $\sigma_2$  &     \multicolumn{1}{c}{a}   &   dof & $\rm \chi^2/dof$ \\
\hline

Mkn\,501  & log10(Flux) &  -10.03$\pm$0.03 & 0.14$\pm$0.02 & 2.26e-10$\pm$3.998e-11 & 6.51e-11$\pm$2.10e-11  & 0.79$\pm$0.09 & 33 & 1.163\\

\hline

Mkn\,421  & log10(Flux) & -10.10$\pm$0.08 & 0.28$\pm$0.08 & 4.07e-10$\pm$5.20e-11 & 2.48e-10$\pm$4.43e-11 &  0.3 & 15 & 0.85\\

\hline

        \end{tabular}
        
 \label{tab:2distpar_lg}
        
\end{table*}


\subsection{Correlation study}

\begin{figure}
\centering
\includegraphics[scale=0.35,angle=-90]{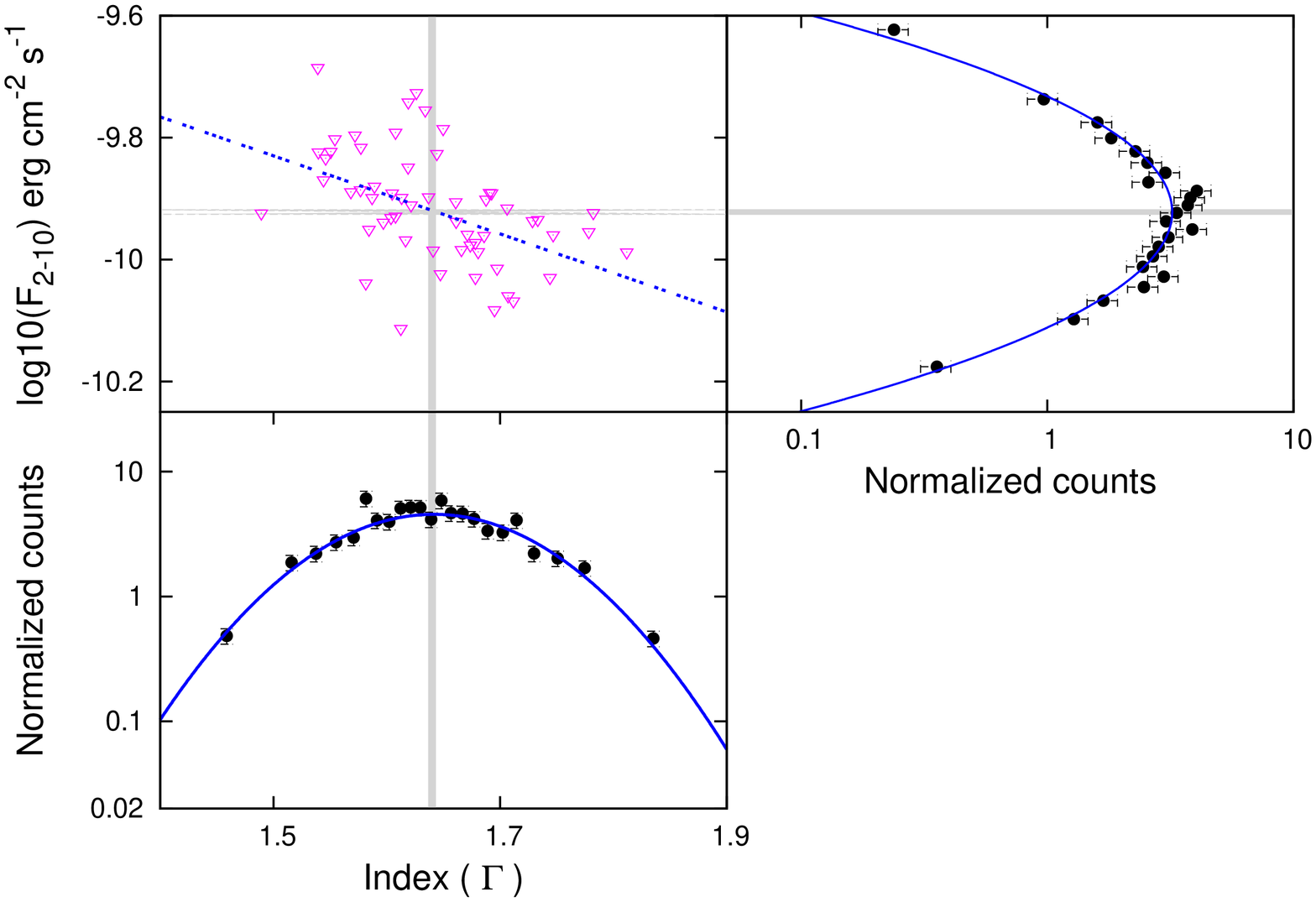}
\caption{Multi-plot for the characterization of flux/index distribution of 3C\,273. Top panel left is the logarithm of flux vs index scatter plot along with the best fit line (dotted line). Top panel right is the histogram of logarithmic of flux distribution. Bottom panel is the histogram of index distribution. The solid curve in the top panel right and bottom panel indicates the best fitted PDF (equation \ref{eq:gauss}). The vertical and horizontal gray bands indicate the 1-$\sigma$ error range on the centroid ($\mu$) of the PDF (equation \ref{eq:gauss}) fitted to the index distribution and  logarithm of flux distribution respectively.}. 
 
\label{fig:273_corr}
\end{figure}

\begin{figure}
\centering
\includegraphics[scale=0.35,angle=-90]{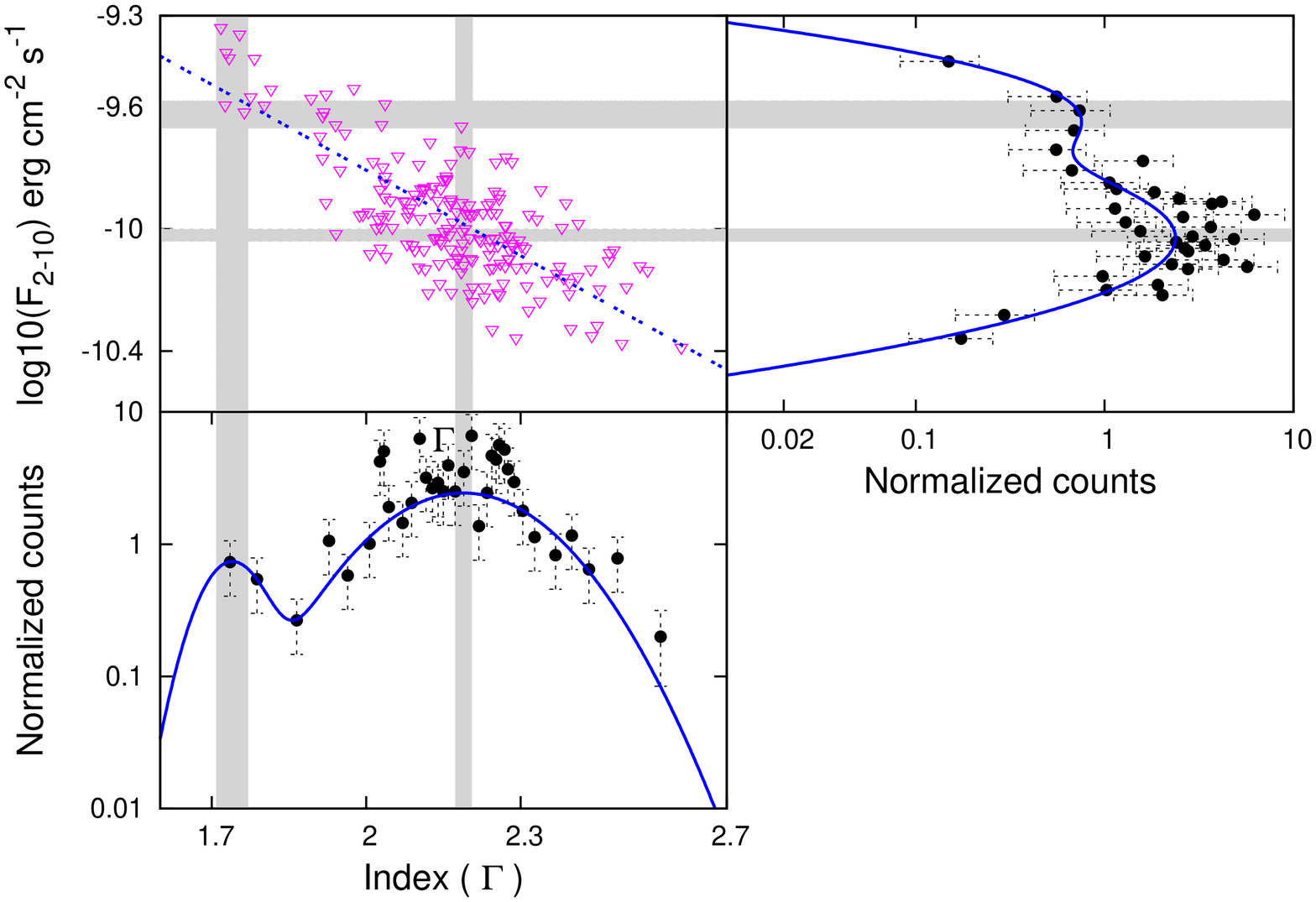}
\caption{Multi-plot for the characterization of flux/index distribution of Mkn\,501. Top panel left is the logarithm of flux vs index scatter plot along with the best fit line (dotted line). Top panel right is the histogram of logarithm of flux distribution. Bottom panel is the histogram of index distribution. The solid curve in the top panel right and bottom panel indicates the best fitted double PDF (equation \ref{eq:dpdf}).  The vertical and horizontal gray bands indicate the 1-$\sigma$ error range on the centroids ($\mu1$, $\mu2$) of the double PDF (equation \ref{eq:dpdf}) fitted to the index distribution and  logarithm of flux distribution respectively.}
\label{fig:501_corr}
\end{figure}

\begin{figure}
\centering
\includegraphics[scale=0.35,angle=-90]{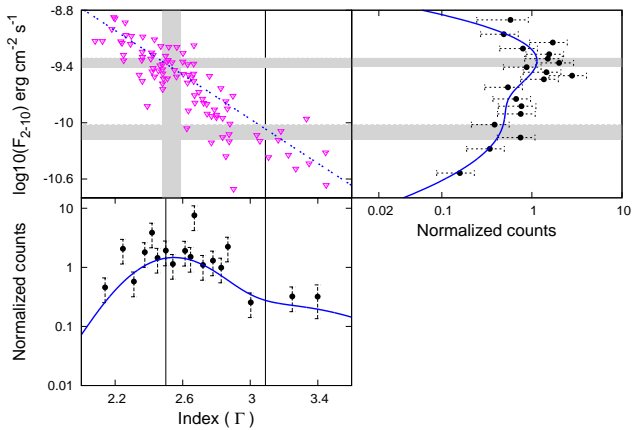}
\caption{Multi-plot for the characterization of flux/index distribution of Mkn\,421. The three panels in this plot are same as described in Fig. \ref{fig:501_corr}. A single vertical line represents the centroid value of the higher index distribution (Section 3.3).}

\label{fig:421_corr}
\end{figure}

The three blazars selected from RXTE catalog has a sufficient number of data points to study the correlation between the photon index and the flux in the 2-10 keV energy band. We have used the Spearman's rank correlation method to assess the correlation behavior between the index and flux. The obtained Spearman's rank correlation coefficient ($r_{s}$), its chance correlation probability (P) and the correlation slope (A) are summarized in Table \ref{tab:corrpar}. The correlation parameters show a significant negative correlation between the flux and index in all the three blazars, which is the usual trend blazars show across the electromagnetic spectrum \citep{2015A&A...573A..69W, 2017ApJ...841..123P, 2011MNRAS.413.2785B}.
In the correlation plots (left top panel in Figs \ref{fig:273_corr}, \ref{fig:501_corr} and \ref{fig:421_corr}), gray bands represent the 1-$\sigma$ error on the centroids of the logarithm of flux and index distributions (Figs \ref{fig:273_corr}, \ref{fig:501_corr} and \ref{fig:421_corr}). In the bottom panel of Fig. \ref{fig:421_corr},  the error on higher index centroid is large, so a single vertical line is shown instead of a gray band.


\begin{table}
\centering
		\caption{ Spearman Correlation results obtained by comparing flux and index distribution of three selected blazars. Col:- 1: Selected blazar sources 2: Spearman's rank correlation coefficient $(r_{s})$, 3: Probability chances for correlation (P) and  4: Slope of the best fitted line to correlation plots (A).}
        \begin{tabular}{ c c c c }
                \hline
                \hline
				Blazar name   &  \multicolumn{1}{c}{$r_{s}$ }          &  \multicolumn{1}{c}{P} &   \multicolumn{1}{c}{A}\\ 
	  
							   \hline

3C\,273 &  -0.60 &  $2.0\times10^{-4}$   & -0.61$\pm$0.22\\   
Mkn\,501 & -0.65 & $2.96\times10^{-24}$ & -0.94$\pm$0.06\\
Mkn\,421 & -0.86 & $7.25\times10^{-29}$ & -1.21$\pm$0.07\\

	 \hline			   
	 \hline    
     
        \end{tabular}
       \label{tab:corrpar}
       						   
\end{table}



\section{Discussion}\label{sect:disc}
After selecting three blazars viz. 3C\,273, Mkn\,501 and Mkn\,421 from the sample of blazars in the RXTE catalog, we used the AD test and histogram fitting to characterize their flux and index distributions.
We found that the flux distribution of FSRQ 3C\,273 follows a log-normal distribution while its index is Gaussian distributed. The log-normal distribution of flux in 3C\,273 is also observed in the monthly binned $\gamma$-ray light curve \citep{2018RAA....18..141S}. Since the variations in the index are related to the fluctuations in acceleration and escape time scales of the emitting particles in the acceleration region, the observed Gaussian distribution in index therefore indicate a linear normal fluctuations in the intrinsic time scales in the acceleration region. This result is consistent with the study by \citealp{2018MNRAS.480L.116S}, where they showed that the linear normal fluctuation in the intrinsic particle acceleration time scale in the acceleration region can produce a log-normal flux distribution. The connection of log-normal flux distribution with the Gaussian index variation indicates that the source of X-ray flux variation in 3C\,273 are not from the accretion disc, instead, the perturbations are mostly local to the jet.

Using the AD test, we found that the flux and index distributions of two BL\,Lacs viz. Mkn\,501 and Mkn\,421 are not consistent with a single distribution, and their histograms can be fitted with a double distribution. Interestingly, the overlapping of the centroids (within 1-$\sigma$ error -- gray bands in Figs \ref{fig:501_corr} and \ref{fig:421_corr}) of the double distributions in the log of flux and index distribution on the best fitted correlation line (blue dotted line in the correlation plots), with same normalization fraction values for both the distributions (Section 3.3, Table \ref{tab:2distpar}), implies that the double distribution in photon index is connected to the double distribution in flux. The reduced $\chi^2$ test shows that the flux distributions of Mkn\,501 and Mkn\,421 are either double log-normal or combination of log-normal and Gaussian, while their index distributions are double Gaussian. However, using the interpretation of \citealp{2018MNRAS.480L.116S}, the double Gaussian distribution in the index would preferably indicate double log-normal distribution in flux. \citealp{2018MNRAS.480L.116S}, showed that Gaussian distribution in the index can be initiated through linear fluctuations in the particle acceleration rate and hence, the log-normal flux distribution may carry information regarding the acceleration processes in the blazar jets. 

In the study of multi-wavelength flux variations in  PKS 1510-089, the two distinct log-normal profiles found in the flux distribution at near-infrared (NIR), optical and $\gamma$-ray energies are connected to two possible flux states in the source \citep{2016ApJ...822L..13K}. In our work, the observation of double Gaussian distribution in index with bi-lognormal flux distribution in Mkn\,501 and Mkn\,421 further confirms the two flux states hypothesis. Moreover, contrary to double log-normal flux distribution in  Mkn\,501 and Mkn\,421 in X-ray, the study of the $\gamma$-ray flux distribution of brightest \emph{Fermi} blazars show a single log-normal flux distribution at $\gamma$-ray energy \citep{2018RAA....18..141S}. However, it should be noted that the results of Shah et al. (2018) are obtained from light curves with a bin size of a month, which is longer than those used in this work.  A longer binning might remove the second component of the distribution. In case of HBL sources like Mkn\,501 and Mkn\,421, the X-ray spectrum lies beyond the synchrotron peak and hence is mainly emitted by the high-energy end of the electron distribution. While in FSRQs, the X-ray emission is mainly due to the low energy tail of the electron distribution. Further, in the case of HBL sources, the low energy $\gamma$-ray spectrum occurs before the break energy of the inverse Compton component. Therefore, the differences observed in the flux distribution of FSRQ 3C\,273 and HBL's (Mkn\,501 and Mkn\,421) at X-rays may possibly be related to the energy of emitting particles. Thus, it seems that low energy emitting particles produce a single log-normal flux distribution while the high energy tail of the electron distribution produces double log-normal flux distribution. However, such inference can be confirmed by carrying out a detailed flux distribution study for a sample of sources with more statistically significant light curves. In this direction, a systematic regular long-term monitoring of blazars with MAXI would be important to probe such information. It will be interesting and important to quantify the flux distribution of these sources in other wave-bands, as that may strengthen the results presented here. However, continuous significant and reliable detection of flux and index at other bands is not available at present. We note that the upcoming Cherenkov Telescope Array \citep{2019EPJWC.20901038C} may provide such high quality light curves in gamma-rays.

\section*{Acknowledgements}
We thank the anonymous referee for valuable comments and suggestions. This work has made use of light curves provided by the University of California, San Diego Center for Astrophysics and Space Sciences, X-ray Group (R.E. Rothschild, A.G. Markowitz, E.S. Rivers, and B.A. McKim), obtained at \url{https://cass.ucsd.edu/~rxteagn/}. R. Khatoon and R. Gogoi would like to thank CSIR, New Delhi (03(1412)/17/EMR-II) for financial support. R. Khatoon would like to thank IUCAA, Pune for the hospitality. R. Gogoi would like to thank IUCAA for associateship.

\bibliographystyle{mnras}
\bibliography{references} 


\bsp	
\label{lastpage}
\end{document}